\documentclass[preprint,showpacs,preprintnumbers,amsmath,amssymb, showkeys,superscriptaddress,titlepage]{revtex4-2}
\usepackage{graphicx}
\usepackage{dcolumn}
\usepackage{bm}
\usepackage{hyperref}
\usepackage{textcomp}

\begin{document}

\title{Highlights of Unstable States in Relativistic Dissociation of Light Nuclei in Nuclear Emulsion}

\author{D.A. Artemenkov}
\affiliation{Joint Institute for Nuclear Research (JINR), Dubna, Russia}

\author{N.K. Kornegrutsa}
\affiliation{Joint Institute for Nuclear Research (JINR), Dubna, Russia}

\author{N.G. Peresadko}
\affiliation{P.N. Lebedev Physical Institute of the Russian Academy of Sciences (LPI), Moscow, Russia}

\author{V.V. Rusakova}
\affiliation{Joint Institute for Nuclear Research (JINR), Dubna, Russia}

\author{A.A. Zaitsev}
\affiliation{Joint Institute for Nuclear Research (JINR), Dubna, Russia}
\affiliation{P.N. Lebedev Physical Institute of the Russian Academy of Sciences (LPI), Moscow, Russia}

\author{P.I. Zarubin}
\email{zarubin@jinr.ru}
\affiliation{Joint Institute for Nuclear Research (JINR), Dubna, Russia}
\affiliation{P.N. Lebedev Physical Institute of the Russian Academy of Sciences (LPI), Moscow, Russia}

\author{I.G. Zarubina}
\affiliation{Joint Institute for Nuclear Research (JINR), Dubna, Russia}

\begin{abstract}
The results of the study of unstable states in relativistic dissociation of isotopes $^{9,7}$Be, $^{10}$B, $^{12,11,10}$C, $^{14}$N and $^{16}$O in nuclear emulsion have been summarized. Their decays are identified in distributions by invariant masses determined by fragment emission angles in the velocity conservation approximation. The observed diversity enables us to assume universality in the formation of nuclear-molecular states near the bond thresholds as a consequence of coalescence of emerging $\alpha$-particles and nucleons.
\end{abstract}


\maketitle

\section{Introduction}

Based on deep traditions and fundamental discoveries, the topic of the relationship between $\alpha$-cluster and nucleon aspects in nuclear structure remains among the paradigms of nuclear physics \cite{1,2,3,4}. Experiments with fast-moving nuclei provide an opportunity to confidently identify of products of their fragmentation and carry out a spectrometry. Progress at accelerators allows this area to remain a dynamically developing section of the Microcosm physics. Advancement into the relativistic region makes an unprecedented diversity of nuclear states fundamentally accessible under conditions of extremely small energy-momentum transfers and, therefore, with minimal perturbation of the initial state. Complementing nuclear clustering, the results in this area are of great value for cosmic ray physics, nuclear astrophysics and in practical applications of beams of relativistic nuclei \cite{5}.

Experiments with relativistic isotopes are mainly aimed at registering heavy fragments with a charge close enough to the charge of the nucleus under study, and as an addition to one or a pair of the lightest nuclei He and H. However, in this an approach, fundamentally important channels containing only fragments of He and H are lost, and, accordingly, the decays of unstable nuclei $^8$Be and $^9$B and more complex states decaying into them.

The opportunity of solving this problem in the nuclear track emulsion motivated the BECQUEREL experiment \cite{6} at the JINR Nuclotron. The method, which retains its uniqueness in the completeness of observations and resolution, is used to study the structure of relativistic fragmentation of an accessible variety of nuclei, including the radioactive ones. Events with the preservation of the primary charge by relativistic fragments are classified as dissociation of the parent nucleus. Among them, there is coherent dissociation not accompanied by fragments of target nuclei (or ``white'' stars) stands out, initiated in the diffraction nuclear and electromagnetic interactions of colliding nuclei. The focus is on ensembles of the lightest nuclei in a narrow angular cone of fragmentation, which remain internally nonrelativistic. Their tracks are fully observed in longitudinally exposed layers of nuclear emulsion from 200 to 500 microns thick. For review, video recordings of dissociation events taken with microscopes have been collected \cite{7}.

The invariant mass of pairs and triplets of He and H fragments is determined from the emission angles in the approximation of conservation of momentum per nucleon of the parent nucleus. In the fragmentation of a number of light nuclei, including the radioactive ones, the decays of $^8$Be(0$^+$), $^{9}$B, $^{12}$C(0$^+_2$) or 
the Hoyle state have been identified as isolated by this parameter in up to several hundred keV (review \cite{8}). A search is underway for the state of $^{16}$O(0$^+_6$) at 660 keV above the 4$\alpha$-threshold with the decay of $\alpha^{12}$C(0$^+_2$) or 2$^8$Be, proposed as a 4$\alpha$-particle Bose-Einstein condensate \cite{3}. 
Extension of this approach to medium and heavy nuclei revealed an enhancement with the number of $\alpha$-particles of the contributions of $^8$Be, $^9$B and $^{12}$C(0$^+_2$), indicating their occurrence during the interaction in the final state \cite{9}.

Of interest is the advancement to higher excitations of light nuclei \cite{10}, which is facilitated by the inertia of the $\alpha$-particle due to the spin pairing of nucleons in it. Another aspect is the expansion of the diversity of fragmentation channels. Many light isotopes have excited states with widths of the order of several eV or lifetimes of several femtoseconds not exceeding approximately 1 MeV above the separation thresholds of the $\alpha$-particle and the stable residue heavier than He \cite{10}. When these states are formed while fragmenting, the decay products will also have minimal scattering angles. They will be an even more convenient subject to study than the $n\alpha$(+$p$)-states. Below we present recent findings made in the analysis of nuclear emulsion exposed by relativistic beams of the isotopes $^9$Be, $^{12}$C, $^{14}$N and $^7$Be.

\section{Dissociation of $^9$B$\textbf{e}$}

The nuclear emulsion is exposed to the beam formed in the fragmentation of $^{10}$B $\to$ $^9$Be nuclei at 2 GeV/$c$ per nucleon \cite{11}. To advance to $^8$Be$^*$ $Q_{2\alpha}$ $>$ 5 MeV excitations, the statistics has been extended to 712 stars $^9$Be $\to$ 2$\alpha$. The search for dissociation $^9$Be $\to$ 2$\alpha$ has been carried out by transverse scanning. The contributions of the decays of $^8$Be(0$^+$) at $Q_{2\alpha} <$ 0.2 MeV, $^8$Be$^*$(2$^+$) at 1 $< Q_{2\alpha} <$ 5 MeV and $^9$Be$^*$(2.43 MeV, $\Gamma$ = 780 eV) \cite{10} at 0.2 $< Q_{2\alpha} <$ 1 MeV \cite{8,12,13} have been established. The leading and close contributions of $^8$Be(0$^+$) and $^8$Be$^*$(2$^+$) indicate a 2-body configuration $n$+($^8$Be(0$^+$)/$^8$Be(2$^+$)) in the ground state of $^9$Be.

The average value of the total transverse momentum in the dissociation $^9$Be $\to$ 2$\alpha$ is about 10 MeV/$c$ per nucleon \cite{12,13}, which is several times less than the Fermi momentum of nucleon motion (100-200 MeV/$c$), which manifests itself in the emission of neutrons. Neglecting the total transverse momentum, one can estimate the transverse momenta carried away by neutrons and, then, estimate the invariant masses $Q_{2\alpha n}$.

Under the condition $Q_{2\alpha}$ $<$ 0.25 MeV in the presence of $^8$Be(0$^+$), a peak appears in the distribution $Q_{2\alpha n}$ (Fig. 1) near the $^8$Be(0$^+$)$p$ threshold equal to 1.665 MeV. Approximation of this distribution by the Breit-Wigner function yields the resonance energy of 1.80 $\pm$ 0.01 MeV at a width $\Gamma$ = 732 keV determined by the experimental resolution and the approximations made. These parameters are consistent with approximately 100 decays of the level $^9$Be$^*$(1.684 MeV, $\Gamma$ = 217 keV, $J^{\pi}$ = $\frac{1}{2}^+$) \cite{10} up to $Q_{2\alpha n}$ $<$ 2 MeV. Their contribution to the $^9$Be $\to$ $^8$Be(0$^+$) channel is 33 $\pm$ 4\% and 14 $\pm$ 2\% to the entire statistics. Under the condition 0.25 $<$ $Q_{2\alpha}$ $<$ 0.85 MeV, i.e. with a veto on $^8$Be(0$^+$) in the region 2 $<$ $Q_{2\alpha n}$ $<$ 3 MeV, where the signal of $^9$Be(2.43) is expected, no more than 25 events or no more than 4\% of the dissociation events $^9$Be $\to$ 2$\alpha$.

\begin{figure}[th]
	\centerline{\includegraphics[width=10cm]{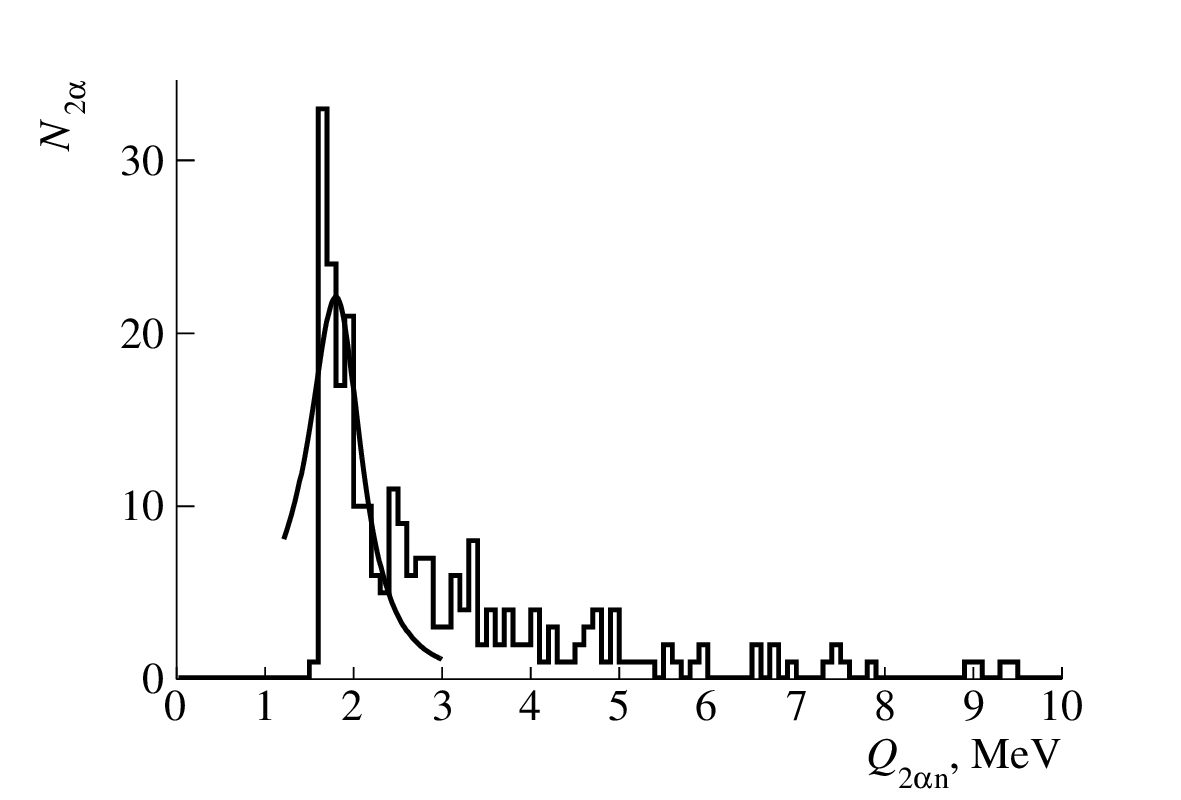}}
	\caption{Distribution over invariant masses of triplets consisting of pairs of $\alpha$-particles $Q_{2\alpha}$ $<$ 0.25 MeV and neutrons in dissociation $^9$Be $\to$  2$\alpha$ at 2 GeV/$c$ per nucleon; curve – Breit-Wigner distribution.}
\end{figure}

The dissociation of $^9$Be $\to$ 2$\alpha$ is a source of $^8$Be$^*$ excitations clearly quantized against the $^9$Be$^*$ background. Among the latter there is $^9$Be (14.4 MeV, $\Gamma$ = 381 eV) \cite{10}, which, like $^9$Be (2.43 MeV), can be reflected in the distribution over $Q_{2\alpha}$. Immediately below the threshold $^7$Li$p$ (17.3 MeV) there is a doublet of levels at 16.6 and 16.9 MeV with the mixed isospin ($\Gamma$ = 108 and 74 keV, $J^{\pi}$ = 2$^+$, T = 0+1) \cite{10} which can be represented as a superposition of 2$\alpha + \alpha$($^3$He$n$/$tp$). Below there is the $^8$Be$^*$(4$^+$) level (11.35 MeV, $\Gamma$ $\approx$ 3.5 MeV). The search for 2$\alpha$-states is available up to 30 MeV, which is facilitated by the dominance of nuclear diffraction hindering the transfer of the angular momentum, and, therefore, the $\alpha$-cluster breakups.

The distribution of $Q_{2\alpha}$($^8$Be) $>$ 5 MeV pairs of $\alpha$-particles indicates their presence throughout the entire range described above (Fig. 2). However, statistics does not allow one to link them with the excitations listed above. Methodological difficulties appear due to the increase in scattering angles, leading to an early exit of tracks from the analyzed emulsion layer. For the same reason, identification by multiple scattering becomes difficult. These data help one to plan experiments with magnetic analysis of relativistic fragments, which become possible while increasing the open angles of $\alpha$-particle pairs.

\begin{figure}[th]
	\centerline{\includegraphics[width=10cm]{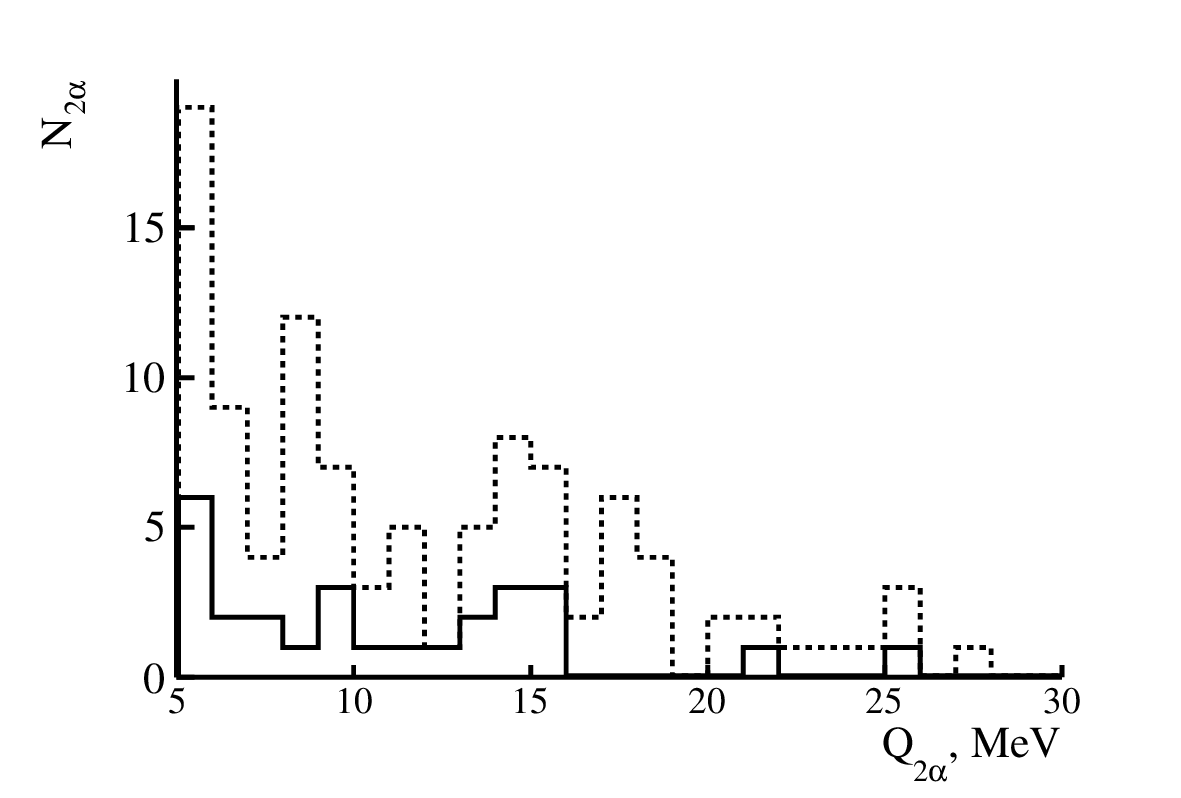}}
	\caption{Distribution over invariant masses $Q_{2\alpha}$ $>$ 5 MeV of $\alpha$-particle pairs in dissociation $^9$Be $\to$ 2$\alpha$ at 2 GeV/$c$ per nucleon (dots), including coherent dissociation (the solid line).}
\end{figure}

\section{Dissociation of $^{12}$C}

Analysis of the invariant mass distributions $Q_{3\alpha}$ of $\alpha$-particle triplets in coherent dissociation $^{12}C$ $\to$ 3$\alpha$ has revealed the decays $^{12}$C(0$^+_2$) \cite{8,14,15}. In the coherent dissociation $^{16}$O $\to$ 4$\alpha$, the enhancement of $^{12}$C(0$^+_2$) and $^8$Be(0$^+$) has been detected. Due to the minimal energy of $^{12}$C(0$^+_2$) and the absence of its contribution in the case of $^{16}$O and other excitations, it is not necessary to introduce the condition on $^{8}$Be(0$^+$).

In the $^{12}$C case, the $Q_{3\alpha}$ distribution contains another peak between 2 and 4 MeV which may correspond to the excitation of $^{12}$C(3$^-$) at 9.64 MeV (or 2.37 MeV above the 3$\alpha$ threshold) with a width of $\Gamma$ = 46 keV \cite{10}. It is interpreted as an equilateral triangular configuration of $\alpha$-particles, each of them has a unit orbital angular momentum with respect to the axis perpendicular to its plane \cite{2}. In the cascade decays of $^{12}$C(3$^-$), as in the case of $^{12}$C(0$^+_2$), the $^8$Be(0$^+$) decays are always present.  The whole family of $\alpha$-particle excitations with energy gaps not more 1 MeV follows \cite{10} above $^{12}$C(3$^-$), below the proton separation threshold (16 MeV).

The emission angles are measured in 510 $^{12}$C $\to$ 3$\alpha$ events with a momentum of 4.5 GeV/$c$ per nucleon. This significant level of statistics is achieved due to a targeted search while transverse scanning. Its part corresponds to a proportional set of 350 coherent dissociations and 160 3$\alpha$-dissociations accompanied with target nuclear fragments. Analysis of the relative yields of $^8$Be(0$^+$), $^{12}$C(0$^+_2$) and $^{12}$C(3$^-$) and the distributions over the transferred transverse momentum of these two types of dissociation in this sample have not revealed differences. The entire statistics is discussed below. The number of $Q_{2\alpha}$($^8$Be) decays $<$ 200 keV in it is equal to 221.

At $Q_{2\alpha}$($^8$Be) $<$ 200 keV, following Fig. 9 in \cite{1}, two peaks are observed in the distribution over $Q_{3\alpha}$ (Fig. 3). The first peak with the average value $Q_{3\alpha}$(RMS) = 417 $\pm$ 27 (165) keV corresponds to $^{12}$C(0$^+_2$), and the second one with the parameters of the Rayleigh distribution $Q_{3\alpha}$($\sigma$) = 2.4 $\pm$ 0.1 MeV – $^{12}$C(3$^-$). Since the peaks are well separated for $^{12}$C(0$^+_2$) decays, the soft condition $Q_{3\alpha}$ $<$ 1 MeV is adopted. The rapid decrease in the contribution $Q_{3\alpha}$ $>$ 4 MeV determines the upper limit of $^{12}$C(3$^-$). The contributions of $^8$Be(0$^+$), $^{12}$C(0$^+_2$) and $^{12}$C(3$^-$) is 221, 57, 97 have been estimated as 43 $\pm$ 4, 11 $\pm$ 2, 19 $\pm$ 2\%, respectively. The contribution of $^{12}$C(0$^+_2$) to the decays of $^8$Be(0$^+$) is equal to 26 $\pm$ 4\%, and $^{12}$C(3$^-$) is 44 $\pm$ 6\% and their ratio is 0.6 $\pm$ 0.1.

\begin{figure}[th]
	\centerline{\includegraphics[width=10cm]{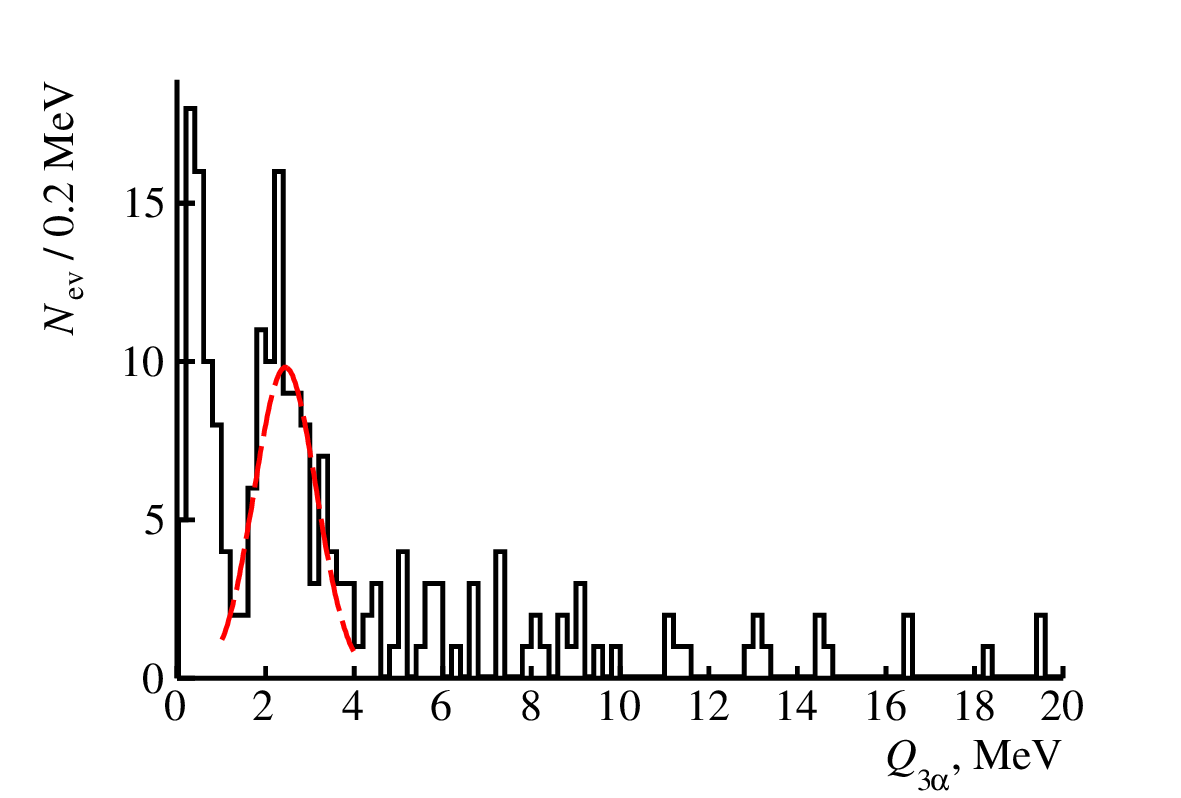}}
	\caption{Distribution over invariant masses $Q_{3\alpha}$ of triplets of $\alpha$-particles in dissociation $^{12}$C $\to$ 3$\alpha$ at 4.5 GeV/$c$ per nucleon with condition $Q_{2\alpha}$($^8$Be) $<$ 200 keV; curve is approximation of $^{12}$C (3$^-$) by Gaussian.}
\end{figure}

\section{Dissociation of $^{14}$N}

The widths of $^8$Be $\Gamma$ = 5.6 eV and $^{12}$C(0$^+_2$) and $\Gamma$ = 9.3 eV \cite{10} result in the fact that relativistic decays occur at distances of the order of several thousand atomic sizes. The $^8$Be(0$^+$)  and $^{12}$C(0$^+_2$) enhancement in the $^{16}$O dissociation indicates that they are special configurations of nuclear matter of extremely low density, and not simply exotic excitations of a part of the parent nucleus. Another argument in favor of this assumption is identification of $^{12}$C(0$^+_2$) in the dissociation of $^{14}$N $\to$ 3$\alpha p$ at 2.9 GeV/$c$ per nucleon \cite{15}. This channel is the leader among peripheral interactions with the transfer of the primary charge to the fragmentation cone \cite{16}. However, the search for the number required to solve this problem is complicated by the diversity of charge states of dissociation of $^{14}$N compared to $^{12}$C \cite{16}, which may include, for example, $^{14}$N $\to$ $^9$B$\alpha$, $^9$Be$^*$(2.43)$\alpha p$, etc.

The $^9$B decays are identified in the emulsion exposed to relativistic $^{10}$B nuclei and fragments $^{12}$C $\to$ $^{10}$C and $^{11}$C \cite{11}. In the coherent dissociation $^{10}$C $\to$ 2$\alpha$2$p$, the condition $Q_{2\alpha p}$($^9$B) $<$ 500 keV and 100\% coincidence of $^8$Be and $^9$B decays has been established \cite{17,18}. In the coherent dissociation of the odd-odd nucleus $^{10}$B $\to$ $^8$Be2$p$, this correspondence decreases to 50 $\pm$ 12\% \cite{19}, and falls in the case of $^{11}$C $\pm$ $^8$Be2$pn$ to 24 $\pm$ 7\% \cite{20}. This behavior can be attributed to the growing contribution of unbound $^9$Be$^*$ excitations which can also suppress $^{12}$C(0$^+_2$).

The $^{14}$N $\to$ 3$\alpha p$ channel provides a coherent picture of the formation of $^8$Be(0$^+$), $^9$B and $^{12}$C(0$^+_2$) \cite{21}. Transverse scanning of the emulsion layers found 226 $^{14}$N $\to$ 3$\alpha p$ stars at 2.9 GeV/$c$ per nucleon, 60 of them have been measured. Among the latter, 30 decays of $Q_{2\alpha}$($^8$Be) $<$ 0.3 MeV have been identified. Ten of these correspond to $^9$B decays ($Q_{2\alpha p}$ $<$ 0.3 MeV), and 6 to $^{12}$C(0$^+_2$) ($Q_{3\alpha}$ $<$ 0.7 MeV). Then, the share of $^8$Be(0$^+$) is 50 $\pm$ 11\%, including $^9$B 16 $\pm$ 6\%  and $^{12}$C(0$^+_2$) decays 10 $\pm$ 4\%. The simultaneous veto on $^9$B and $^{12}$C(0$^+_2$) leaves 13 $^8$Be(0$^+$) decays and 8 $\alpha$-pairs in the region of $^9$Be$^*$(2.43) 0.3 $<$ $Q_{3\alpha}$ $<$ 0.7 MeV (Fig. 4). The estimate of the contribution of $^9$Be$^*$(2.43) is 13 $\pm$ 6\%. Reflecting the identification, these data reflect the status of the measurements.

\begin{figure}[th]
	\centerline{\includegraphics[width=7cm]{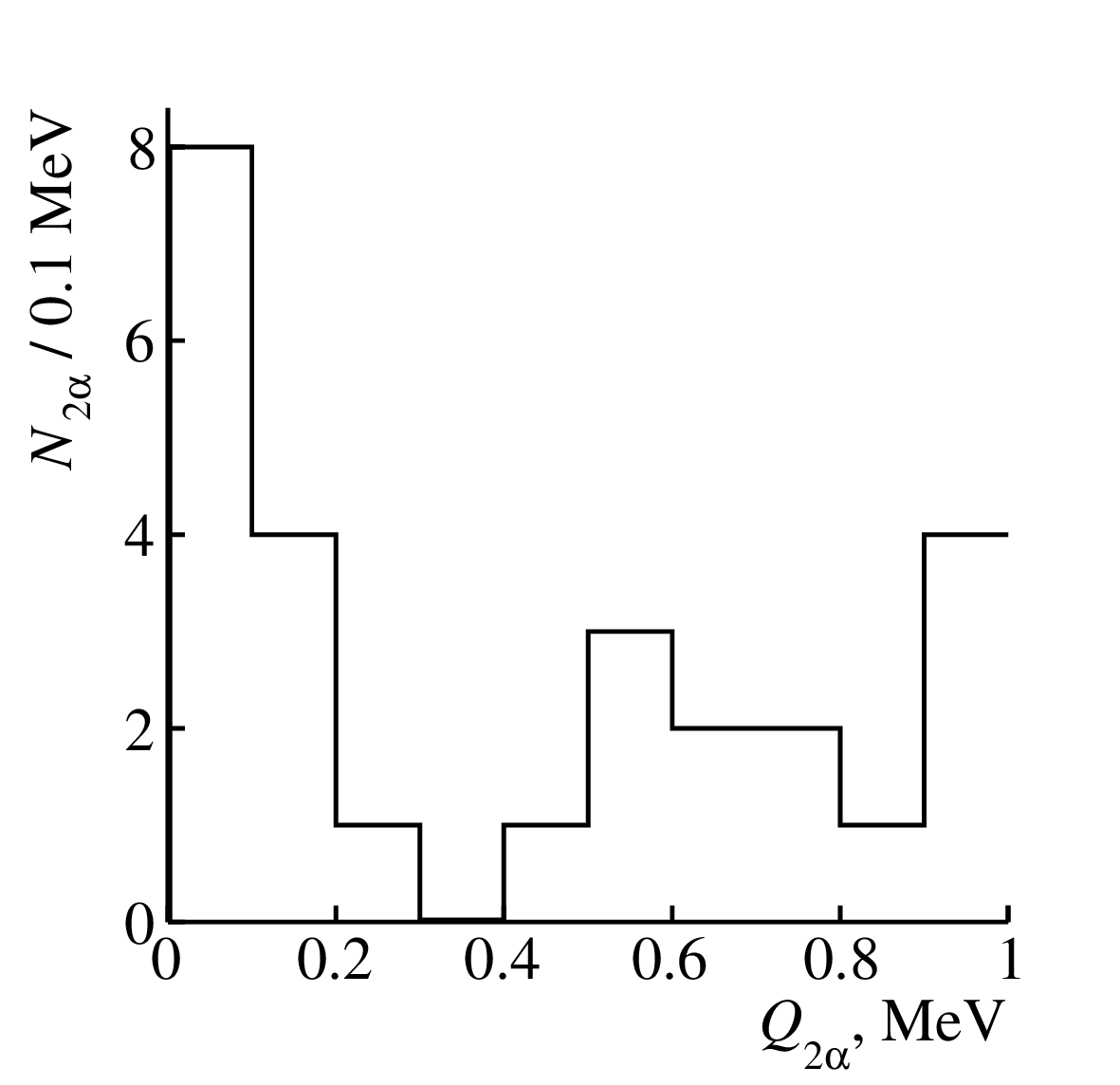}}
	\caption{Distribution over invariant masses $Q_{2\alpha}$ of $\alpha$-particle pairs in the dissociation $^{14}$N $\to$ 3$\alpha p$ at 2.9 GeV/$c$ per nucleon with veto on $^{12}$C(0$^+_2$) ($Q_{3\alpha}$ $>$ 0.7 MeV) and $^9$B ($Q_{2\alpha p}$ $>$ 0.5 MeV).}
\end{figure}

\section{Dissociation of $^{7}$B$\textbf{e}$}

Previously, the dissociation of $^7$Be nuclei at the momentum of 1.8 GeV/$c$ per nucleon was investigated in \cite{22,23}. The $^7$Be beam was formed both in the charge exchange $^7$Li $\pm$ $^7$Be, and in the composition of fragments of close magnetic rigidity $^{12}$C $\to$ $^{12}$N, $^{10}$C, $^7$Be \cite{11}. Having an independent significance with respect to the $^4$He-$^3$He clustering. These data are required in the contribution of $^7$Be including its excitations to the dissociation of $^8$B, $^{9,11}$C, $^{12}$N \cite{5}.

Among the dissociation events, the decay $^6$Be $\to$ $^4$He + 2$p$ has been identified on the $Q_{\alpha 2p}$ invariant masses in the $^7$Be $\to$ $^6$Be + $n$ events \cite{15}. They accounted for about 27\% in the $^7$Be $\to$ $^4$He + 2$^1$H channel. In addition, several $^7$Be $\to$ $^6$Li + $p$ events were observed. Despite the small statistics, the distribution over the invariant mass $Q_{Li+p}$ allows one to identify 5 events near the 5.6 MeV threshold. Their average value, equal to 6.4 $\pm$ 0.3 at RMS 0.6, corresponds to the $^7$Be(7.2) level with $\Gamma$ = 0.40 keV \cite{10}. Despite the absence of the isospin prohibition, the small width of this level happens due to the difference in its configuration from $^4$He-$^3$He.

\section{Conclusions and Prospects}

The presented results on the study of the unstable states in relativistic dissociation of light isotopes indicate the unique productivity of the nuclear emulsion method. Determination of the invariant masses of $\alpha$-particle ensembles from the fragment emission angles in the approximation of conservation of momentum per nucleon of the parent nucleus (or its velocity) allows one to identify the decays of $^8$Be(0$^+$), $^8$Be(2$^+$), $^9$Be(1.7), $^9$B, $^6$Be, $^{12}$C(0$^+_2$), $^{12}$C(3$^-$) and $^7$Be(7.2). The decisive factor is the record spatial resolution and sensitivity of the nuclear emulsion. For light stable nuclei, the assumption that singly and doubly charged fragments correspond to the isotopes $^1$H and $^4$He turns out to be sufficient.

The diversity of the states listed above allows one to assume a common nature of the emergence of nuclear-molecular states immediately above the bond thresholds in the coalescence of already emerged real $\alpha$-particles and nucleons. The time and cross-sections of the interaction of fragments moving with low relative velocities in the system of the parent nucleus or their ensembles may be sufficient for the lowest-energy synthesis reactions. The justification of this scenario requires to  include low-energy nuclear physics data for to describe of the relativistic nuclear fragmentation.

Because of the widest coverage of nucleus energy (from several hundred MeV to tens of GeV per nucleon) and the uncertainty of the centers of mass in the cone of relativistic fragmentation, there is a problem of uniform comparison of the data obtained in nuclear emulsion and modeling. It is resolved in the relativistic-invariant approach proposed by A. M. Baldin to describe multiple particle collisions in relativistic nucleus interactions in the space of relative 4-velocities \cite{24,25}. Actually, the analysis presented above is a particular implementation for the situation of extremely small squares of the difference in the 4-velocities of particles (or Lorentz factors of relative motion near one). An essential point in it is the certainty in the mass of the ensemble participants.

Formation of $^8$Be(0$^+$), $^9$B, $^{12}$C(0$^+_2)$, $^9$Be(1.7) indicates that conditions for nuclear astrophysics processes can be reproduced in the cone of relativistic fragmentation. On the other hand, the presented observations can serve as ``beacons'' which provide confidence in searching for complex compositions of nuclear matter at the lower limit of nuclear temperature and density. Fundamentally, there are no unsolvable problems along this path, except, of course, the labor intensity of direct analysis and the cost of the emulsion itself, where the proportion of suitable dissociation events is small. Hopefully, the progress in intelligent microscopy and image recognition will allow us to expand the scope of studies of multiple nuclear ensembles at the binding thresholds based on the nuclear emulsion method.


\begin{thebibliography}{0}
\bibitem{1}  A.H. Wuosmaa,  R.R.Betts,  M.Freer, and  B.R. Fulton, \href{https://doi.org/10.1146/annurev.ns.45.120195.000513}{\textit{Ann. Rev. Nucl. Part. Sci.} \textbf{45}, 89 (1995).} 

\bibitem{2} M. Freer, \href{https://link.springer.com/book/10.1007/978-3-642-45141-6}{\textit{Lect. Notes in Phys.} \textbf{879} (2014) Springer Int. Publ.}

\bibitem{3} A. Tohsaki, H. Horiuchi, P. Schuck, and G. Ropke, \href{https://doi.org/10.1103/RevModPhys.89.011002}{\textit{Rev. Mod. Phys.} \textbf{89}, p.011002 (2017).} 

\bibitem{4} I. Lombardo, D. Dell’Aquila, \href{https://doi.org/10.1007/s40766-023-00047-4}{\textit{La Rivista del Nuovo Cimento} \textbf{46}, 521 (2023).} 

\bibitem{5}  P.I. Zarubin, \href{https://doi.org/10.1007/978-3-319-01077-9_3}{\textit{Clusters in Nuclei Volume 3} (2014) 51.} \href{https://arxiv.org/1309.4881}{arXiv: 1309.4881.} 

\bibitem{6} \href{http://becquerel.jinr.ru/}{The BECQUEREL Project web-site.} 

\bibitem{7} \href{http://becquerel.jinr.ru/movies/movies.html}{The BECQUEREL Project web-site, “movies” page.}  

\bibitem{8} D.A. Artemenkov \textit{et al.}, \href{https://doi.org/10.1140/epja/s10050-020-00252-3}{\textit{Eur. Phys. J. A} \textbf{56}, 250 (2020).}
 
\bibitem{9}  A.A. Zaitsev, D.A. Artemenkov, V.V. Glagolev, M.M. Chernyavsky, N.G. Peresadko, V.V. Rusakova, P.I. Zarubin, \href{DOI 10.1016/j.physletb.2021.136460}{\textit{Phys. Let. B} \textbf{820} (2021) 136460.} \href{https://arxiv.org/2102.09541}{arXiv:2102.09541.} 

\bibitem{10} F. Ajzenberg-Selove, \href{http://www.tunl.duke.edu/NuclData/}{\textit{Nucl. Phys. A} \textbf{490} (1988).}

\bibitem{11} P.A. Rukoyatkin \textit{et al.}, \href{https://doi.org/10.1134/S1063778807070149}{\textit{Eur. Phys. J.} \textbf{162} (2008) 267.} \href{https://arxiv.org/pdf/1210.1540}{arXiv:1210.1540.} 

\bibitem{12} D.A. Artemenkov \textit{et al.}, \href{https://link.springer.com/article/10.1134/S1063778807070125}{\textit{Phys. Atom. Nucl.} \textbf{70} (2007) 1222.} \href{https://arxiv.org/pdf/nucl-ex/0605018}{arXiv: nucl-ex/0605018.}

\bibitem{13} D.A. Artemenkov \textit{et al.}, \href{https://doi.org/10.1007/s00601-008-0307-6}{\textit{Few Body Syst.} \textbf{44} (2008) 273.}

\bibitem{14} D.A. Artemenkov \textit{et al.}, \href{https://doi.org/10.1016/j.radmeas.2018.11.005}{\textit{Rad. Meas.} \textbf{119} (2018) 199.} \href{https://arxiv.org/abs/1812.09096}{arXiv:1812.09096.}

\bibitem{15} D.A. Artemenkov \textit{et al.}, \href{https://doi.org/10. 1007/978-3-030-32357-8_24}{\textit{Springer Proc. Phys.} \textbf{238} (2020) 137.}

\bibitem{16}  T. V. Shchedrina \textit{et al.}, \href{https://doi.org/10.1134/S1063778807070149}{\textit{Phys. Atom. Nucl.} \textbf{70} (2007) 1230.} \href{http://arxiv.org/abs/nucl-ex/0605022}{arXiv:nucl-ex/0605022.}

\bibitem{17} D.A. Artemenkov \textit{et al.}, \href{https://doi.org/10.1007/s00601-011-0223-z}{\textit{Few Body Syst.} \textbf{50} (2011) 259.}
 
\bibitem{18} K.Z. Mamatkulov \textit{et al.}, \href{https://doi.org/10.1134/S1063778813100141}{\textit{Phys. Atom. Nucl.} \textbf{76} (2013) 1224.} \href{ http://arxiv.org/abs/1309.4241v1}{arXiv:1309.4241.}

\bibitem{19} A.A. Zaitsev \textit{et al.}, \href{https://link.springer.com/article/10.1134/S1063779617060612}{\textit{Phys. Part. Nuclei} \textbf{48} (2017) 960.}

\bibitem{20} D.A. Artemenkov, A.A. Zaitsev and P.I. Zarubin, \href{https://doi.org/10.1134/S1063779617010026}{\textit{Phys. Part. Nucl.} \textbf{48} (2017) 147.} \href{https://arxiv.org/abs/}{arXiv:1607.08020.} 

\bibitem{21} E. Mitsova \textit{et al.}, \href{https://doi.org/10.1134/S1063779622020575}{\textit{Phys. Part. Nucl.} \textbf{53} (2022) 456.} \href{https://arxiv.org/pdf/2011.06265}{arXiv:2011.06265.} 

\bibitem{22} N.G. Peresadko \textit{et al.}, \href{https://doi.org/10.1134/S1063778807070137 }{\textit{Phys. Atom. Nucl.} \textbf{70} (2007) 1226.} \href{https://arxiv.org/pdf/nucl-ex/0605014}{arXiv:nucl-ex/0605014.} 

\bibitem{23} N.K. Kornegrutsa  \textit{et al.}, \href{https://doi.org/10.1007/s00601-014-0832-4}{\textit{Few Body Syst.} \textbf{55} (2014) 1021.} \href{https://arxiv.org/pdf/1410.5162}{arXiv:1410.5162.} 

\bibitem{24} A.M. Baldin, L.A. Didenko, V.G. Grishin, A.A. Kuznetsov, Z.V. Metreveli, \textit{Zeitchrift fur Physik C.} \href{https://doi.org/10.1007/BF01552542}{\textbf{33} (1987) 363.}

\bibitem{25} Baldin A. M. and Didenko L. A., \href{https://doi.org/10.1002/prop.2190380402}{\textit{Fortschritte der Physik/Progress of Physics} \textbf{38} (1990) 261.}

\end{thebibliography}
\end{document}